\documentclass[%
reprint,
superscriptaddress,
bibnotes,
 amsmath,amssymb,
 aps,
]{revtex4-2}

\usepackage{graphicx}
\usepackage{dcolumn}
\usepackage{bm}
\usepackage{caption}
\usepackage{subcaption}
\usepackage{hyperref}

\begin{document}


\title{ First direct measurement of $^{59}$Cu(p,$\alpha$)$^{56}$Ni:\\
A step towards constraining the Ni-Cu cycle in the Cosmos\\}

\author{J.S. Randhawa}\email{jrandhaw@nd.edu}
 \affiliation{Department of Physics, University of Notre Dame, Notre Dame, IN 46556}
 \affiliation{Joint Institute for Nuclear Astrophysics, University of Notre Dame, Notre Dame, Indiana 46556, USA}
 
\author{R. Kanungo}%

\affiliation{Astronomy and Physics Department, Saint Mary’s University,
 Halifax, Nova Scotia B3H 3C3, Canada}%
 \affiliation{TRIUMF, Vancouver, BC V6T2A3, Canada}

\author{J. Refsgaard}
\affiliation{Astronomy and Physics Department, Saint Mary’s University,
 Halifax, Nova Scotia B3H 3C3, Canada}%
\affiliation{TRIUMF, Vancouver, BC V6T2A3, Canada}

\author{P. Mohr}
\affiliation{Institute for Nuclear Research (Atomki), P.O. Box 51, Debrecen H-4001, Hungary}
\author{T. Ahn}
\affiliation{Department of Physics, University of Notre Dame, Notre Dame, IN 46556}
 \affiliation{Joint Institute for Nuclear Astrophysics, University of Notre Dame, Notre Dame, Indiana 46556, USA}

\author{M. Alcorta}
\affiliation{TRIUMF, Vancouver, BC V6T2A3, Canada}
\author{C. Andreoiu}
\affiliation{Department of Chemistry, Simon Fraser University, Burnaby, British Columbia V5A 1S6, Canada}
\author{S. S. Bhattacharjee}
\affiliation{Astronomy and Physics Department, Saint Mary’s University,
 Halifax, Nova Scotia B3H 3C3, Canada}%
\affiliation{TRIUMF, Vancouver, BC V6T2A3, Canada}
\author{B. Davids}
\affiliation{TRIUMF, Vancouver, BC V6T2A3, Canada}
\affiliation{Department of Physics, Simon Fraser University, Burnaby, British Columbia V5A 1S6, Canada}

\author{G. Christian}
\affiliation{Astronomy and Physics Department, Saint Mary’s University,
 Halifax, Nova Scotia B3H 3C3, Canada}
 
\author{A. A. Chen }
\affiliation{Department of Physics and Astronomy, McMaster University, Hamilton, Ontario L8S 4M1, Canada}

\author{R. Coleman}
\affiliation{Department of Physics, University of Guelph, Guelph, Ontario N1G 2W1, Canada}

\author{P. Garrett}
\affiliation{Department of Physics, University of Guelph, Guelph, Ontario N1G 2W1, Canada}

\author{G. F. Grinyer}
\author{E. Gyabeng Fuakye}
\affiliation{Department of Physics, University of Regina, Regina, SK S4S 0A2, Canada}
\author{G. Hackman}
\affiliation{TRIUMF, Vancouver, BC V6T2A3, Canada}

\author{R. Jain}
\affiliation{Facility for Rare Isotope Beams, Michigan, 48824, U.S.}
\author{K. Kapoor}
\affiliation{Department of Physics, University of Regina, Regina, SK S4S 0A2, Canada}
\author{R. Kr\"ucken}
\affiliation{TRIUMF, Vancouver, BC V6T2A3, Canada}
\affiliation{Department of Physics and Astronomy, University of British Columbia, Vancouver, BC V6T 1Z1, Canada}

\author{A. Laffoley}
\affiliation{Department of Physics, University of Guelph, Guelph, Ontario N1G 2W1, Canada}
\author{A. Lennarz}
\affiliation{TRIUMF, Vancouver, BC V6T2A3, Canada}
\affiliation{Department of Physics and Astronomy, McMaster University, Hamilton, Ontario L8S 4M1, Canada}
\author{J. Liang}
\affiliation{Department of Physics and Astronomy, McMaster University, Hamilton, Ontario L8S 4M1, Canada}

\author{Z. Meisel}
\affiliation{Institute of Nuclear and Particle Physics, Department of Physics \& Astronomy, Ohio University, Athens, OH 45701, USA}

\author{N. Nikhil}
\affiliation{Astronomy and Physics Department, Saint Mary’s University,\\
 Halifax, Nova Scotia B3H 3C3, Canada}%
 
\author{A. Psaltis}
\affiliation{Department of Physics and Astronomy, McMaster University, Hamilton, Ontario L8S 4M1, Canada}
\author{A. Radich}
\author{M. Rocchini}
\affiliation{Department of Physics, University of Guelph, Guelph, Ontario N1G 2W1, Canada}
\author{N. Saei} 
\affiliation{Department of Physics, University of Regina, Regina, SK S4S 0A2, Canada}

\author{M. Saxena}
\affiliation{Institute of Nuclear and Particle Physics, Department of Physics \& Astronomy, Ohio University, Athens, OH 45701, USA}

\author{M. Singh}
\affiliation{Astronomy and Physics Department, Saint Mary’s University,
 Halifax, Nova Scotia B3H 3C3, Canada}
 
\author{C. Svensson}
\affiliation{Department of Physics, University of Guelph, Guelph, Ontario N1G 2W1, Canada}

\author{P. Subramaniam}
\affiliation{Astronomy and Physics Department, Saint Mary’s University,
 Halifax, Nova Scotia B3H 3C3, Canada}

\author{A. Talebitaher}
\affiliation{Department of Physics, University of Regina, Regina, SK S4S 0A2, Canada}
 
\author{S. Upadhyayula}
\affiliation{TRIUMF, Vancouver, BC V6T2A3, Canada}

\author{C. Waterfield}
\affiliation{Astronomy and Physics Department, Saint Mary’s University,
 Halifax, Nova Scotia B3H 3C3, Canada}
 \author{J. Williams}
\author{M. Williams}
\affiliation{TRIUMF, Vancouver, BC V6T2A3, Canada}


\begin{abstract}
Reactions on the proton-rich nuclides drive the nucleosynthesis in Core Collapse Supernovae (CCSNe) and in X-ray bursts (XRBs). CCSNe eject the nucleosynthesis products to the interstellar medium and hence are a potential inventory of p-nuclei, whereas in XRBs nucleosynthesis powers the light curves. In both astrophysical sites the  Ni-Cu cycle, which features a competition between $^{59}$Cu(p,$\alpha$)$^{56}$Ni and $^{59}$Cu(p,$\gamma$)$^{60}$Zn, could potentially halt the production of heavier elements. Here, we report the first direct measurement of $^{59}$Cu(p,$\alpha$)$^{56}$Ni using a re-accelerated $^{59}$Cu beam and cryogenic solid hydrogen target. Our results show that the reaction proceeds predominantly to the ground state of $^{56}$Ni and the experimental rate has been found to be lower than Hauser-Feshbach based statistical model predictions. New results hints that the $\nu p$-process could operate at higher temperatures than previously inferred and therefore remains a viable site for synthesizing the heavier elements.

\end{abstract}

\maketitle

In the Universe most of the heavy elements, not made in the slow neutron-capture in stellar burning, are produced via rapid neutron capture (r-process) proposed to occur in neutron star mergers (NSMs)\cite{Kasen17,Smartt17}. The recent discovery of gravitational waves from  NSMs and follow-up multi-wavelength observations have bolstered the NSMs as a viable site for heavy elements synthesis\cite{Smartt17}. However, there are several nuclides ($\sim$ 30 nuclides of 23 elements) that cannot be synthesized in the $r$-process or $s$-process. Especially, the mechanism for the production of the light $p$-nuclei, $^{92,94}$Mo and $^{96,98}$Ru, is still debatable\cite{Bliss_2018, Meyer, Goriely}. Nucleosynthesis on the proton-rich side, e.g. $\nu p$-process in Core-Collapse SuperNovae (CCSNe) and $rp$-process in type-I X-ray bursts (XRBs) has been suggested as sites where these p-nuclei can be synthesized\cite{Carla, Arcones_2011, Schatz2001}. In both $rp$-process and $\nu p$-process, as soon as the reaction flow reaches $^{59}$Cu, the $^{59}$Cu(p,$\alpha$) and $^{59}$Cu(p,$\gamma$) reactions start competing due to the lower $\alpha$-emission threshold in $^{60}$Zn compared to the proton threshold. This leads to the Ni-Cu cycle occurring in two different astrophysical sites i.e. in the X-ray bursts ($rp$-process) and in CCSNe ($\nu p$-process)\cite{Wormer,Arcones2012, Cyburt2016}. $^{59}$Cu(p,$\alpha$)$^{56}$Ni  returns the cycle to $^{56}$Ni, while $^{59}$Cu(p,$\gamma$) breaks out of the Ni-Cu cycle and takes the flow further, depending on the (p,$\gamma$)/(p,$\alpha$) rate ratio. In the case of the $\nu p$-process, if $^{59}$Cu(p,$\alpha$)$^{56}$Ni is dominating over (p,$\gamma$) over the wide range of relevant temperatures, there is little flow above $^{59}$Cu and hence the $\nu p$-process cannot be a contender for the synthesis of heavier p-nuclei. As for $rp$-process, the ashes of XRBs do not become part of the interstellar medium and they are therefore an unlike source of heavy nuclei. Instead, they are buried deeper in the neutron star which plays an important role in determining the thermal profile of neutron star crust. However, the Ni-Cu cycle significantly affects the energy generation and hence the shape of XRB light curves. Hence $^{59}$Cu(p,$\alpha$)$^{56}$Ni is one of the few identified reactions which directly impacts the XRB light curves and hinders the XRB light curve model-observation comparison\cite{Cyburt2016}. Therefore, it is of foremost importance to measure $^{59}$Cu(p,$\alpha$)$^{56}$Ni in addition to $^{59}$Cu(p,$\gamma$)$^{60}$Zn to understand the Ni-Cu cycle in the $\nu p$ process and in XRBs.\\

In this work, we focus on the $^{59}$Cu(p,$\alpha$)$^{56}$Ni reaction. Currently, there is no experimental information on this reaction rate. Relevant temperature range for XRBs and $\nu p$-process is $\sim$1GK and 1-4GK, respectively. The corresponding Gamow window is 1.1-1.4 MeV for XRBs and 1.1 to 4.04 MeV for $\nu p$-process. Direct measurement of $^{59}$Cu(p,$\alpha$)$^{56}$Ni in the Gamow window is an arduous task because the predicted cross-sections are very small and production of high intensity radioactive $^{59}$Cu beam is very challenging. Therefore, in an alternative approach there have been attempts to measure the time-inverse reaction cross-sections i.e. $^{56}$Ni($\alpha$,p)$^{59}$Cu \cite{Proposal_Konrad}. However, these time-inverse measurements are valid if $^{59}$Cu(p,$\alpha$)$^{56}$Ni exclusively proceeds to the ground state of $^{56}$Ni. Current estimates of $^{59}$Cu(p,$\alpha$)$^{56}$Ni and time-inverse $^{56}$Ni($\alpha$,p) are based on the Hauser-Feshbach based statistical model codes. However, the validity needs to be ascertained against a direct measurement of $^{59}$Cu(p,$\alpha$)$^{56}$Ni reaction cross section. A few recent experiments including $^{33}$Cl(p,$\alpha$)$^{30}$S \cite{Deibel_2011}, $^{34}$Ar($\alpha$,p)$^{37}$K\cite{Long2017} have provided first hints of large discrepancies between experimental data and predicted (p,$\alpha$) and ($\alpha$,p) reaction rates on neutron deficient nuclei. Most importantly, above mentioned cases have a similar level density in the compound nucleus as expected for $^{60}$Zn. It is known that $\alpha$-optical potentials perform poorly compared to experiments\cite{Avri17}. Therefore, it is important to perform a  direct measurement of $^{59}$Cu(p,$\alpha$)$^{56}$Ni at energies above the Gamow window where cross sections are higher. The results can be used to test the validity of the Hauser-Feshbach approach commonly used to predict the stellar $^{59}$Cu(p,$\alpha$) rate and its inverse i.e. $^{56}$Ni($\alpha$,p)$^{59}$Cu, and to constrain Hauser-Feshbach model parameters.\\

We report the first direct measurement of $^{59}$Cu(p,$\alpha$)$^{56}$Ni using the IRIS facility with cryogenic solid H$_{2}$ target and re-accelerated $^{59}$Cu beam at TRIUMF. We  provide the  total cross-sections at a center-of-mass energy (E$_{c.m.}$)$=$6.0 MeV and demonstrate that the significant contribution comes from populating the ground state of $^{56}{\rm Ni}$.\\

\begin{figure}
    \centering
    \includegraphics[width=\linewidth]{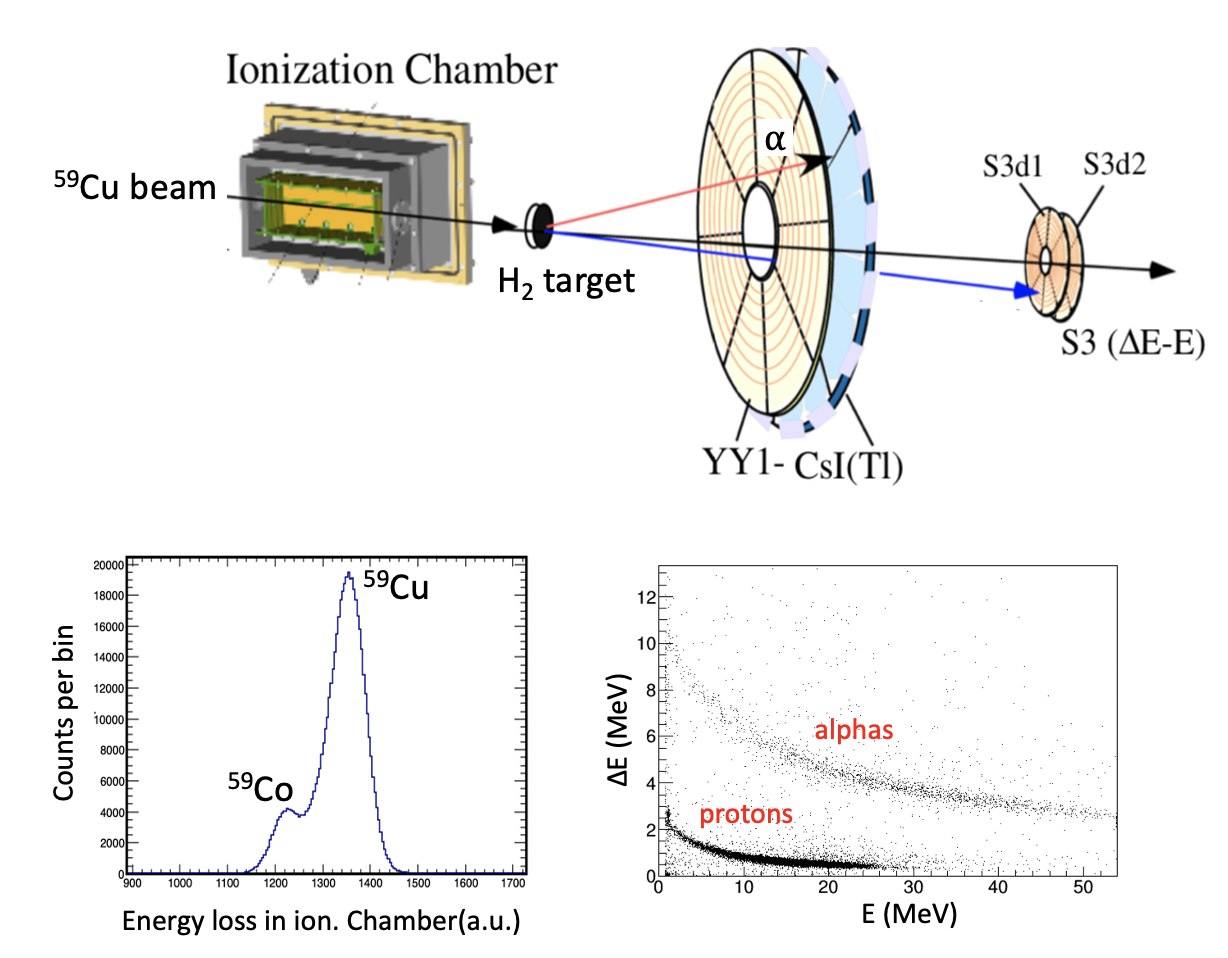}
   
    \caption{The upper panel shows the schematic of IRIS set-up which includes an ionization chamber followed by a cryogenic solid H$_{2}$ target and two $\Delta$E-E telescopes for particle identification. The bottom left panel shows the energy loss spectrum of beam particles in the ionization chamber and the bottom right panel shows the particle identification using the $\Delta$E-E telescope (i.e. YY1-CsI(Tl).}
    \label{fig1}
\end{figure}

\begin{figure}
    \centering
    \includegraphics[width =\linewidth]{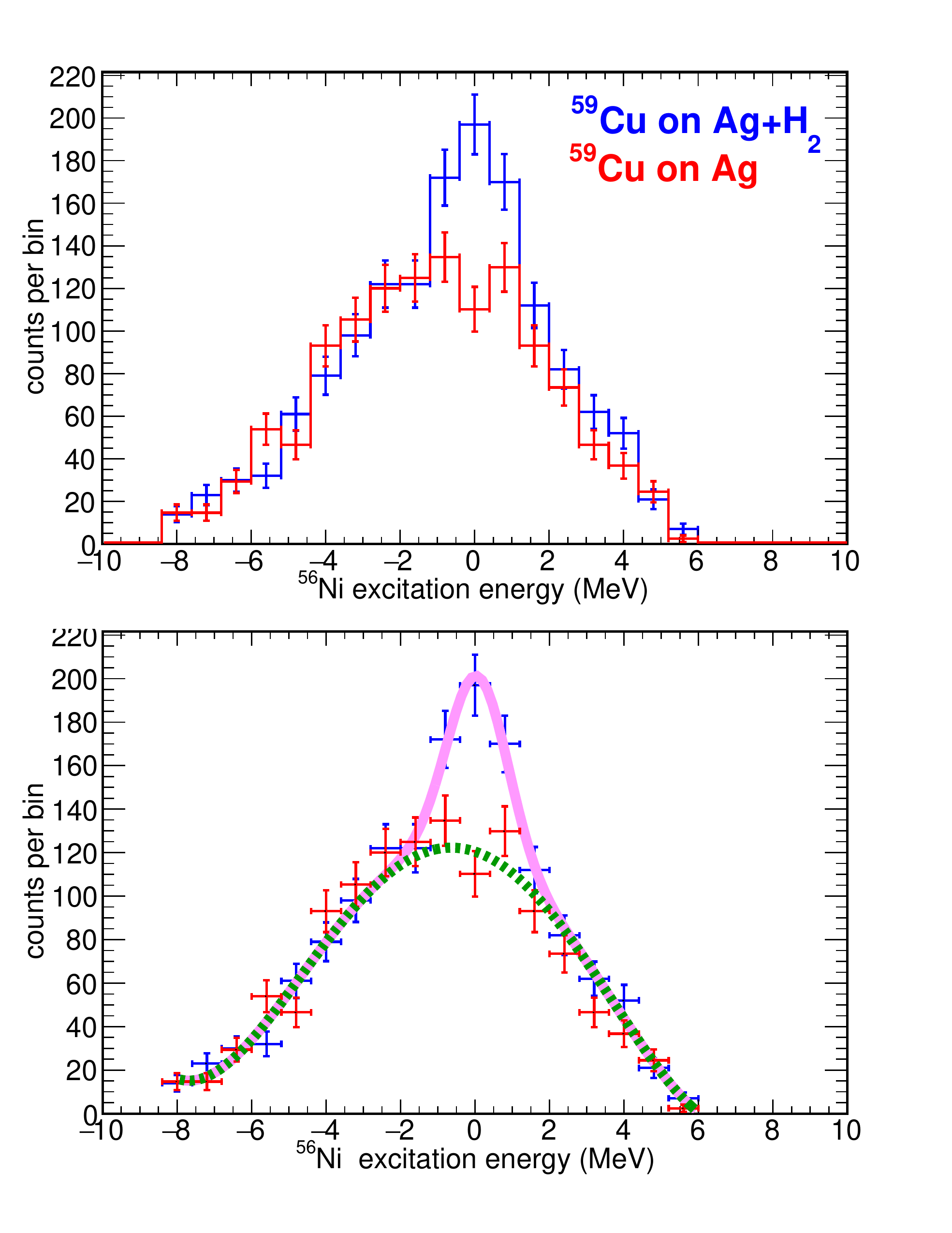}
    \caption{The upper panel shows the excitation energy spectrum with (blue) and without (red) H$_{2}$ target. The blue histogram shows a clear ground state peak above the background (red histogram). The lower panel shows the blue histogram fitted with a Gaussian+polynomial function (solid pink line) and background fit (polynomial only) is also shown by the green dotted line. }
    \label{fig2}
\end{figure}
\begin{figure}
    \centering
    \includegraphics[width =\linewidth]{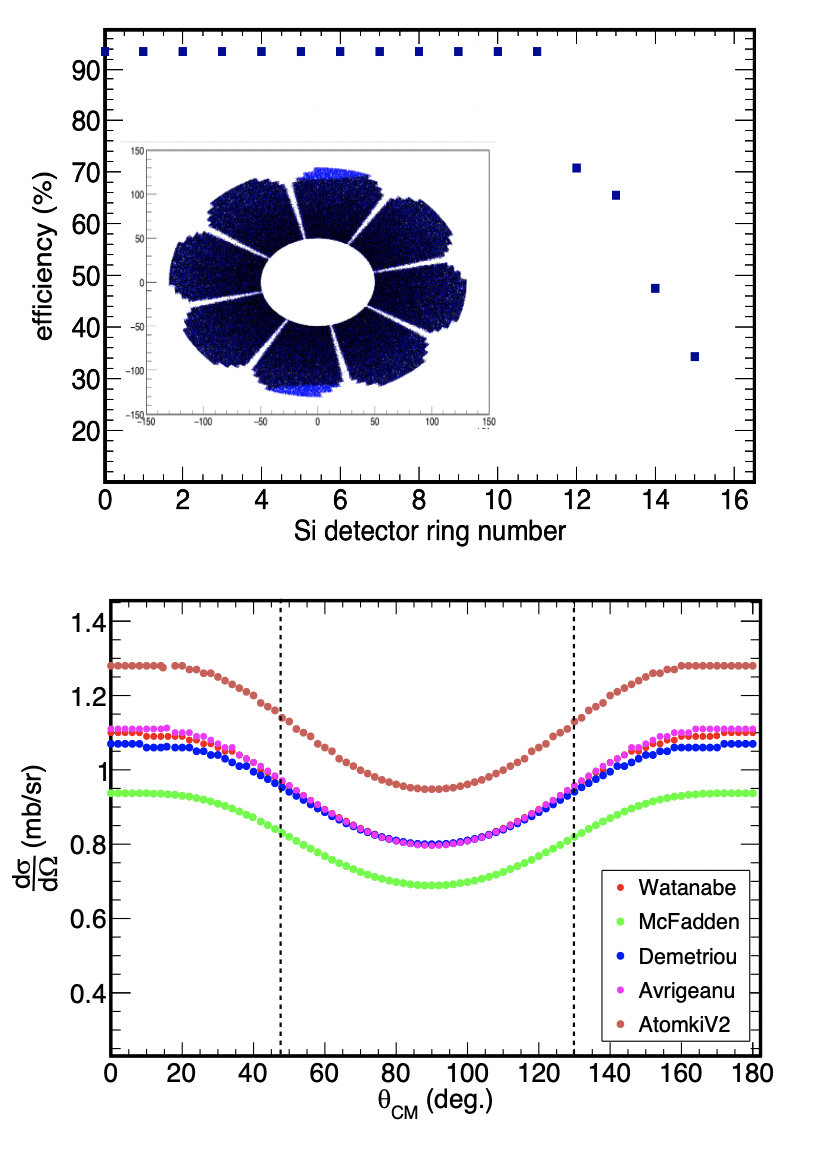}
    \caption{The upper panel shows the simulated detector efficiency as a function of ring number (which provides angle) taking into account the masking of part of a detector by heat shield around the detector. Inset shows the simulated hit pattern with heat shield. The lower panel shows the calculated angular distribution using different $\alpha$-OMPs in TALYS.  Vertical dotted lines show the $\theta_{CM}$ coverage corresponding the angles covered in laboratory in this experiment.}
    \label{fig3}
\end{figure}

\begin{figure}
    \centering
    \includegraphics[width =\linewidth]{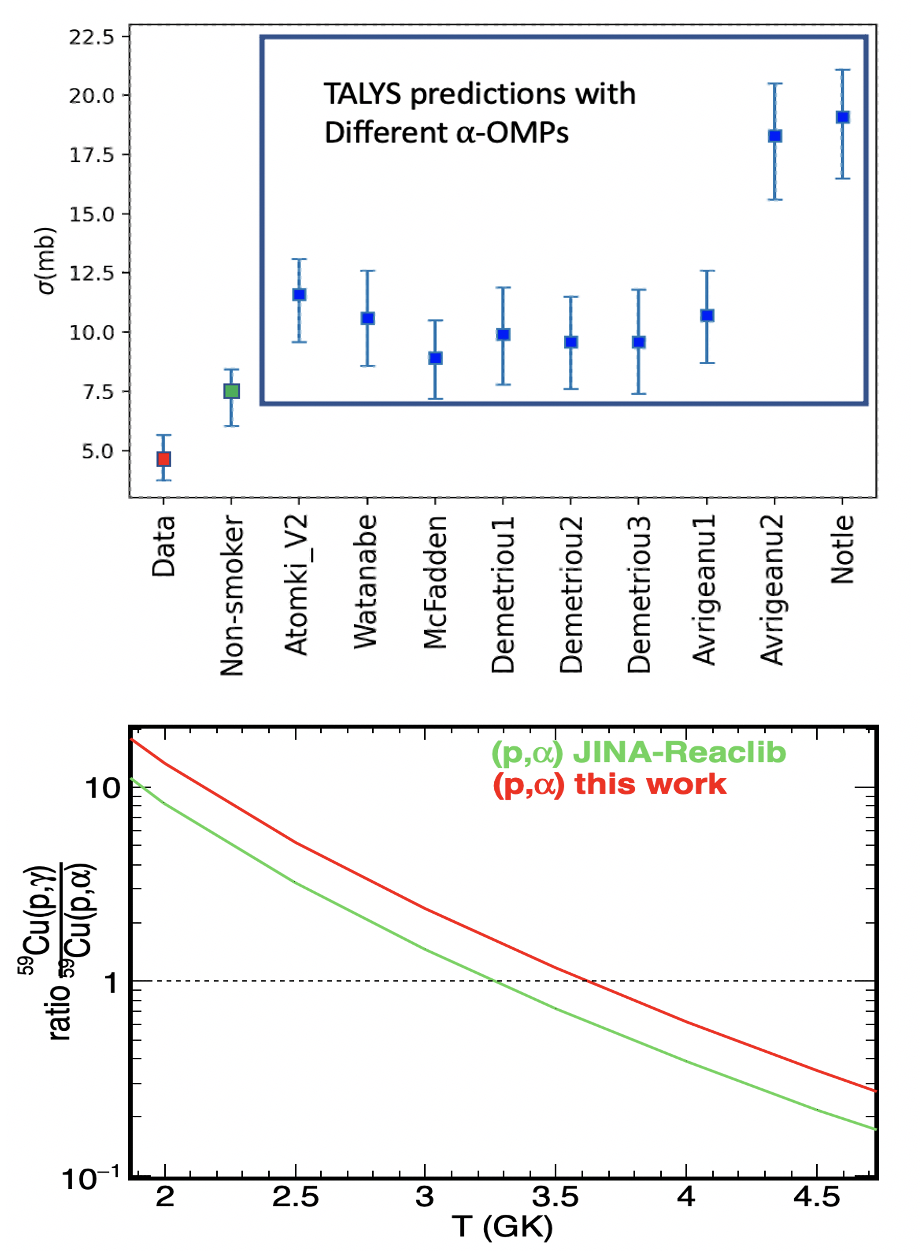}
    \caption{The top panel shows the cross-section obtained in the current work compared to various statistical model calculations at E$_{C.M.} =$6.0 MeV. The new cross-section is lower compared to all the statistical model predictions. Error bars on the calculated cross-sections reflect total change in cross-sections inside the target. The lower panel shows the ratio of $^{59}$Cu(p,$\gamma$) to $^{59}$Cu(p,$\alpha$) with default (p,$\alpha$) rate in JINA-Reaclib (i.e. non-smoker based) and with the new (p,$\alpha$) reaction rate.}
    \label{fig:fig4}
\end{figure}


\textit{Experiment details.} The experiment was performed using IRIS facility in ISAC-II at TRIUMF. A schematic of the detector layout of IRIS is shown in Fig.~\ref{fig1} and for more details please see ref.\cite{Kanungo2014}. The radioactive beam of $^{59}$Cu was produced via spallation of a Niobium target with 480 MeV protons. The $^{59}$Cu beam was re-accelerated using the ISAC-II superconducting LINAC to 8.5$A$ MeV and then passed through an ionization chamber, filled with isobutane gas at 19.5 Torr at room temperature. The average beam intensity was $\sim$ 3600 pps. The energy loss of the beam measured in this ionization chamber provided an event-by-event identification of the $^{59}$Cu incident beam and its contaminant $^{59}$Co throughout the experiment. Following this, the beam interacts with a thin windowless solid hydrogen (H$_{2}$) reaction target built on a 4.3 $\mu$m thick Ag foil backing facing upstream of the H$_{2}$ layer. The target cell with the foil was cooled to $\sim$4$^{\circ}$K before forming solid H$_{2}$. The energy of the elastically scattered beam on the Ag foil was measured with and without H$_{2}$, providing continuous measurement of the target thickness during the experiment. These scattered beam particles were detected using a double-sided silicon strip detector placed 52.5 cm downstream of the target, covering laboratory angles of 1.2$^{\circ}$-3.8$^{\circ}$. The average H$_{2}$ target thickness was 53$\mu$m, and the target thickness between the first and last run of data-taking period showed a change of 7$\%$ over the entire data taking period. Protons and $\alpha$-particle from reactions were detected using annular arrays of 100 $\mu$m thick single-sided silicon strip detectors followed by a layer of 12 mm thick CsI(Tl) detectors placed 15cm downstream of the target. This detector combination served as an energy-loss and total energy (E) telescope for identifying the p and $\alpha$ recoils after the target. The CsI(Tl) detectors were calibrated using $^{59}$Cu(p,p)$^{59}$Cu elastic scattering. The detector telescope covered scattering angles of $\theta_{lab}$ = 18.5$^{\circ}$–40.7$^{\circ}$.\\

\textit{Results.}
The excitation energy spectrum of $^{56}$Ni, shown in Fig.~\ref{fig2}(upper panel), was reconstructed using the missing mass technique using the energy and scattering angle of the $\alpha$-particles, measured by the silicon-CsI(Tl) ($\Delta E$-E) telescope. The narrow peak centered around $~\sim$0 MeV in the excitation energy spectrum is the ground state of $^{56}$Ni. Energy of first excited state in $^{56}$Ni is 2.7 MeV and hence easily resolved from the ground state in current experiment. One of the major sources of background is $\alpha$-particles originating from the reactions on the Ag foil. The background from the Ag foil was measured by collecting data without H$_{2}$ target and is shown with a red dashed-dotted histogram (Fig.~\ref{fig2} lower panel) normalized by the incident beam intensity. In this experiment, $\theta_{lab}$ = 18.5$^{\circ}$–35$^{\circ}$ where lower angle comes from experimental coverage and higher angle is maximum allowed angle of $\alpha$-particle at this energy. This corresponds to $\theta_{c.m.}$= 48$^{\circ}$-130$^{\circ}$ when accounting for the experimental acceptance in angle as well as energy. Fig.~\ref{fig3} (upper panel) shows the detection efficiency using Monte-Carlo simulations. Due to a heat shield surrounding the solid hydrogen target, the efficiency drops at higher angles, and obtained spectra were corrected for this efficiency. The total cross section for $^{59}$Cu(p,$\alpha$)$^{56}$Ni corresponds to integration from $\theta_{c.m.}$=0$^{\circ}$-180$^{\circ}$. Since our experimental coverage of $\theta_{c.m.}$ is limited, therefore, in order to get angle-integrated counts, the angular distribution was calculated using code TALYS \cite{DEMETRIOU2002,Avri17}. Angular distributions obtained using different $\alpha$-optical model potentials ($\alpha$-OMPs) are shown in Fig.~\ref{fig3}. The ratio of integrated cross-section in the experimental acceptance to the total cross-section provides the correction factor of 0.62 to the experimental results and variation in this correction factor, using angular distribution from different potentials, provides an estimate of the systematic uncertainty. The center-of-mass energy (E$_{c.m.}$) at the beginning and the end of the solid H$_{2}$ target is 5.7 MeV and 6.3 MeV, respectively. E$_{c.m.}$ at the center of target corresponds to 6.0 MeV whereas weighted energy, defined as $\int \sigma(E)E dE/\int \sigma(E)dE$, is 6.02 MeV(where the energy dependence of HF based $\tt{NON-SMOKER}$ cross sections was used). Therefore, in this work cross-sections are provided at E$_{c.m.}=$6.0 MeV 

From the excitation energy spectrum, a major highlight is that $^{59}$Cu(p,$\alpha$)$^{56}$Ni, within current measurement sensitivity,  proceeds exclusively to the ground state of $^{56}$Ni. Hence, at the center-of-mass energy(E$_{c.m.}$)$=$6.0 MeV, $^{59}$Cu(p,$\alpha$)$^{56}$Ni$_{g.s}$ is equal to the total $^{59}$Cu(p,$\alpha$)$^{56}$Ni cross-section. The measured cross-section at this energy is shown in Fig.~\ref{fig:fig4} (in the top panel, red square). Experimental error bars reflect both statistical and systematic uncertainties. The systematic uncertainty contains 5\% contribution from the beam counts, 5\% from target thickness, and 15\% from angular distribution and 10\% from simulated detection efficiency. Fig.~\ref{fig:fig4} upper panel shows the comparison of the experimental cross-section to statistical model calculations which includes results from the Non-Smoker (database)\cite{RAUSCHER20001} and TALYS using various input $\alpha$-Optical Model Potentials($\alpha$-OMPs)\cite{MCFADDEN1966, DEMETRIOU2002,Avri2014,Avri94, Notle,Mohr20}. Other options used in TALYS calculations are, the phenomenological proton-OMP   and constant temperature Fermi gas model for level densities(i.e ldmodel 1). Error bars on calculated(theoretical) cross-sections reflects the change in cross-section across the H$_{2}$ target (i.e. accounts for total cross-section change inside the target).The experimental cross-section is lower compared to all the Hauser-Feshbach-based statistical model predictions. 
In general, the (p,$\alpha$) cross section in the statistical model depends on the transmissions $T_i$ in the entrance and exit channels. Very schematically,
\begin{equation}
\sigma(p,\alpha) \sim \frac{ T_{p,0} T_\alpha} { \sum_i T_i}
\end{equation}
where at the experimental energy the sum in the denominator is dominated by the elastic and inelastic proton channels. Thus, $\sigma(p,\alpha)$ is essentially sensitive only to the chosen $\alpha$-OMP whereas other ingredients of the statistical model like the nucleon-OMP, the gamma-ray strength function, and the level density have only marginal influence. Interestingly, all recent $\alpha$-OMPs predict $(p,\alpha)$ cross sections around 10 mb, thus overestimating the experimental result by about a factor of two. A somewhat smaller deviation is found for the McFadden/Satchler $\alpha$-OMP (Fig.4, top panel).\\
\\
\textit{Impact on $\nu$p-process and XRBs:}
In the work of \citet{Arcones2012}, it was shown that  that the $\nu p$-process starts to efficiently produce heavy elements only when the temperature drops below $\sim$3 GK. At higher temperatures,  the reaction $^{59}$Cu(p,$\alpha$)$^{56}$Ni is faster than the reaction $^{59}$Cu(p,$\gamma$)$^{60}$Zn and hence cycle the reaction flow back to $^{56}$Ni. To understand the impact of the measured cross section on the flow of $\nu$p-process, one needs to compare the $^{59}$Cu(p,$\gamma$)$^{60}$Zn with $^{59}$Cu(p,$\alpha$)$^{56}$Ni reaction rate. Currently, these rates are based on the statistical model (in JINA-REACLIB \cite{Cyburt_2010}).   We obtained the new $^{59}$Cu(p,$\alpha$) reaction rate, at different temperatures, assuming the energy dependence from the $\tt{NON-SMOKER}$ reaction rate and scaled it down by a factor of 1.6 based on our measured value. Fig.~\ref{fig:fig4} (lower panel) shows the ratio of (p,$\gamma$) to (p,$\alpha$) reaction rates using  $^{59}$Cu(p,$\alpha$)$^{56}$Ni $\tt{NON-SMOKER}$ reaction rate of JINA-Reaclib  and new $^{59}$Cu(p,$\alpha$)$^{56}$Ni reaction rate obtained in this work. This plot shows that the (p,$\alpha$) reaction starts exceeding the (p,$\gamma$) at temperatures higher than 3.7 GK. Hence, the Ni-Cu cycle will become effective at higher temperatures in the $\nu$p-process than previously inferred. This implies that flow in $\nu$p-process can proceed to higher mass regions over a wider range of temperatures making the production of $^{92,94}$Mo and $^{96,98}$Ru p-nuclei and other nuclei viable, provided other conditions are conducive. This conclusion is based on an assumption that at lower temperatures, scaled down reaction rate follows the energy dependence of Non-smoker rate. However, this conclusion will still be valid even if the reaction cross section within the Gamow window is a factor of two higher or lower than shown in Fig.4 lower panel. A factor of two lower reaction cross-section means the (p,$\alpha$) reaction will only start dominating over (p,$\gamma$) reaction even at temperatures higher than 3.7 GK, whereas for a factor of two higher reaction cross-section, $\nu p$-process will still synthesize the heavier nuclides at temperature below $\sim$3 GK, hence upholds the conclusions drawn in reference\cite{Arcones2012}.\\
\\
However, the situation for XRBs is more complex, where temperatures of interest are 1 GK or below i.e. lower than that of $\nu$p-process. A recent measurement of nuclear level density in $^{60}$Zn shows an unexpected plateau at the energies relevant for XRBs\cite{Soltesz}. It remains to be seen whether the statistical model is valid in the temperature range of XRBs or not. Therefore, for the XRBs, further measurements are required to understand the contribution of individual resonances to both $^{59}$Cu(p,$\alpha$)$^{56}$Ni and $^{59}$Cu(p,$\gamma$)$^{60}$Zn reaction rates. The current experiment shows that $^{59}$Cu(p,$\alpha$)$^{56}$Ni predominantly proceeds to the ground state of $^{56}$Ni, therefore, time-inverse reaction i.e $^{56}$Ni($\alpha$,p)$^{59}$Cu measurement could be a viable option too as more intense $^{56}$Ni beams are possible compared to the challenging production of $^{59}$Cu beam. Nonetheless, either direct or time-inverse measurements are required in the XRB Gamow window to infer the applicability of statistical models and  will help elucidate the role of the Ni-Cu cycle in XRBs. \\

To summarize, we report the first direct measurement of $^{59}$Cu(p,$\alpha$)$^{56}$Ni reaction cross-section using a pure solid H$_{2}$ target at the IRIS facility at TRIUMF. The new measurement shows that $^{59}$Cu(p,$\alpha$)$^{56}$Ni proceeds predominantly to the ground state of $^{56}$Ni. The new cross-section is a factor of 1.6 to 4 lower compared to commonly used statistical model predictions. The new reaction rate when compared to $^{59}$Cu(p,$\gamma$)$^{60}$Zn shows that the (p,$\alpha$) reaction  starts dominating at temperatures above 3.7 GK, higher than previously inferred by theoretical model based reaction rates, favoring the synthesis of heavier nuclides in the $\nu p$-process.

\begin{acknowledgments}
We would like to acknowledge the support of the beam delivery group at TRIUMF. JSR  thanks A. Simon for useful discussion. J.S.R and T.A. were supported by NSF grant no. 2011890. The support from NSERC, CFI and Research Nova Scotia are gratefully acknowledged. ZM is supported by U.S. Department of Energy, Office of Science grant No. DE-FG02-88ER40387 and DE-SC0019042. TRIUMF receives funding via a contribution through the National Research Council Canada. The support from RCNP for the target is gratefully acknowledged. It was partly supported by the grant-in-aid program of the Japanese government under the contract number 23224008 and 14J03935. 

\end{acknowledgments}


\bibliography{apssamp}

\end{document}